\documentstyle[sprocl]{article}
\newcommand{\PSbox}[3]{\mbox{\rule{0in}{#3}\includegraphics{#1}\hspace{#2}}}
\def\be{\begin{equation}}
\def\ee{\end{equation}}
\def\bea{\begin{eqnarray}}
\def\eea{\end{eqnarray}}

\begin{document}
\vspace*{-2cm}

hep-ph/9611309 \hfill MIT-CTP-2586

\vspace*{0.5cm}

\title{ARE THE SUPERSYMMETRIC HIGGS PARTICLES PSEUDO-GOLDSTONE
BOSONS?~\footnotemark[1]}
\author{CSABA CS\'AKI}
\address{Center for Theoretical Physics \\ 
Massachusetts Institute of Technology \\ 
Cambridge, MA 02139, USA}
\maketitle\abstracts{
A promising solution to the doublet-triplet splitting problem
of SUSY GUT's is the Higgs as pseudo-Goldstone boson
mechanism. We present two models which naturally implement
this idea and extend one of them to include fermion
masses. We examine the phenomenological consequences of
this mechanism and present the favored parameter region.}
\footnotetext[1]{Based on work done in collaboration
with Lisa Randall and Zurab Berezhiani.}

The perhaps most problematic aspect of SUSY GUT's is the 
doublet-triplet splitting problem. The problem is to 
explain the large separation in mass scales between 
the Higgs doublet and triplet fields without introducing 
fine-tuning~\cite{Lisatalk}. The original motivation for considering SUSY 
theories was to eliminate the fine-tuning needed to keep
the Higgs doublets light. Therefore it would be embarrassing 
to reintroduce such a fine-tuning into the theory.

The only solution based on a symmetry principle
to the doublet-triplet splitting problem 
(and therefore, perhaps, the most natural one)
is the Higgs as pseudo-Goldstone boson mechanism.
In this picture the explanation for why the Higgs doublets
are light is that they are pseudo-Goldstone bosons  
of a spontaneously broken global accidental symmetry of the
Higgs sector~\cite{Jap}. When Yukawa couplings (couplings of the
Higgs sector to matter fields) are incorporated, the accidental
global symmetry is explicitly broken. However, because of the
non-renormalization theorems the Higgs mass can only be of the
order of the SUSY-breaking, or weak scale. 

The only known implementation of this mechanism that can be made
natural is based on the $SU(6)$ gauge group~\cite{Zurab}.
In this model the accidental symmetry of the superpotential
arises because there are two sectors of the chiral superfields
responsible for gauge symmetry breaking that do not mix and thus
the global symmetry of that sector is $SU(6)\times SU(6)$.
During spontaneous symmetry breaking one of the global
$SU(6)$'s breaks to $SU(4)\times SU(2)\times U(1)$, while the
other to $SU(5)$. The diagonal (gauged) $SU(6)$ thus breaks
to $SU(3)\times SU(2)\times U(1)$. There are exactly
two light doublets in this model, 
so after adding the matter fields
the low-energy particle content is that of the MSSM. This 
accidental global $SU(6)\times SU(6)$ symmetry could be enforced
by a discrete symmetry that forbids the mixing of the two
sectors of the Higgs fields. This symmetry however has to be 
such, that even higher order mixing terms suppressed by the ratio
$M_{GUT}/M_{Pl} \sim 10^{-3}$ are forbidden up to
$(M_{GUT}/M_{Pl})^5$ not to give large masses to the
Higgs doublets. It is very difficult to find explicit 
realizations for such a model. The reason is that usually the more
one suppresses the terms breaking the accidental global symmetry
the more fine tuning is needed to get the right values of VEV's 
from the superpotential. Thus one would again need a small parameter
in the Lagrangian. Two possible solutions to this problem have
been presented in Ref.~\cite{BCR}. Here we present the simpler 
of the two models. This model will make use of the small parameters
that have to be present in the theory anyway: the soft breaking
terms.

Consider an $SU(6)\times Z_n$ GUT theory where the Higgs 
sector consists of an adjoint field $\Sigma$ and a fundamental
and antifundamental field $H+\bar{H}$. If we assume that 
$(\bar{H}H)$
has charge $n$ under $Z_n$ while $\Sigma$ is invariant, then the 
most general superpotential allowed by the gauge and discrete
symmetries is given by 
\begin{equation}
W(\Sigma ,H,\bar{H})=\frac{1}{2}M{\rm Tr}\Sigma^2+
\frac{1}{3}\lambda {\rm Tr}\Sigma^3 +\frac{\alpha}{M_{Pl}^{2n-3}}
(\bar{H}H)^n.
\end{equation}
After the addition of the soft SUSY breaking terms to the scalar
potential, the VEV's are determined by
$\langle \Sigma \rangle ={\rm diag}(1,1,1,1,-2,-2)$,
$\langle H \rangle=\langle \bar{H} \rangle=a 
(\frac{M_{weak}}{M_{Pl}})^{\frac{1}{2n-2}} M_{Pl}$,
where the coefficient $a$ depends on the soft breaking
terms. 
The first allowed term mixing the two sectors is
$\frac{1}{M_{Pl}^{2n-2}}(\bar{H}H)^{2n-2}(\bar{H}\Sigma H)$,
whose contribution to the Higgs mass is acceptably small.
One can modify this theory such that one of the
sectors includes two adjoints $\Sigma_1$ and $\Sigma_2$ and
the fields $\bar{H},H$ in the other sector, and the 
superpotential ($n\geq 4$)
\begin{equation}
W(\Sigma_1,\Sigma_2 ,H,\bar{H})=M{\rm Tr}\Sigma_1\Sigma_2+
\frac{1}{3}\lambda_1 {\rm Tr}\Sigma_1^3 
+\frac{1}{3}\lambda_2 {\rm Tr}\Sigma_2^3
+\frac{\alpha}{M_{Pl}^{2n-3}} (\bar{H}H)^n,
\end{equation}
and an additional discrete $Z_3$ symmetry under which 
$Q_{\Sigma_1}=-Q_{\Sigma_2}=\frac{1}{3},Q_{\bar{H}H}=0$. 
One can extend this model to include matter fields as well. 
We add the $SU(6)$ representations~\cite{Bar} $(15+\bar{6}+\bar{6}')_i+20$,
where $i=1,2,3$. A discrete $Z_n\times Z_n\times Z_3$
symmetry can enforce a realistic Yukawa coupling structure.
The light fields are in exact correspondence with the
fields of the MSSM, and
the top Yukawa coupling is of ${\cal O}(1)$, because this
is the only coupling of the light fields arising from a
renormalizable term. The hierarchy in the fermion masses
arises naturally due to the suppression of the
nonrenormalizable terms by $\langle H\rangle /M_{Pl}\sim 1/30$,
$\langle \Sigma \rangle /M_{Pl}\sim 1/1000$. This is therefore
a complete SUSY GUT model, which solves both the doublet-triplet
splitting problem and the fermion hierarchy problem~\cite{BCR}.

The phenomenology of this (and similar) models can be tested
using the fact that in these models the $\mu$-term  at the 
GUT-scale is fixed
to be~\cite{Jap} $m_0^2+\mu^2=-B\mu$, where $m_0$ is the common soft 
breaking scalar mass at the GUT scale, and $B$ is the soft 
breaking parameter corresponding to the $\mu$-term. This means
that the number of independent MSSM parameters in this model 
is reduced by one. After taking the RGE running between 
the GUT and the weak scale and the requirement of radiative
symmetry breaking into account, one obtains an equation~\cite{CR}
for $\tan \beta$ in terms of the other input parameters
$(m_0,M_{1/2},A_0)$ and the top Yukawa coupling $\lambda_t$.
This equation does not have a solution for every values of the
input parameters, thus restricting the parameter space
of this model. It is interesting to note, that the equation
for $\tan \beta$ is not symmetric under $\tan \beta\to -\tan
\beta$, because the above mentioned boundary condition
breaks this symmetry. A typical plot for the
favored parameter region where the equation for
$\tan \beta$ has a solution is given in Fig.~1, together
with the values $\tan \beta$ can take on in this regime~\cite{CR}.
Thus the assumptions on the GUT-scale Higgs sector physics result in
testable predictions for weak-scale physics. 

In summary, we have presented a complete SUSY GUT model which
solves the doublet-triplet splitting problem, the $\mu$-problem
and the fermion hierarchy problem  and we have discussed the
implications of this model for weak-scale physics.

\begin{figure}
\begin{center}
\PSbox{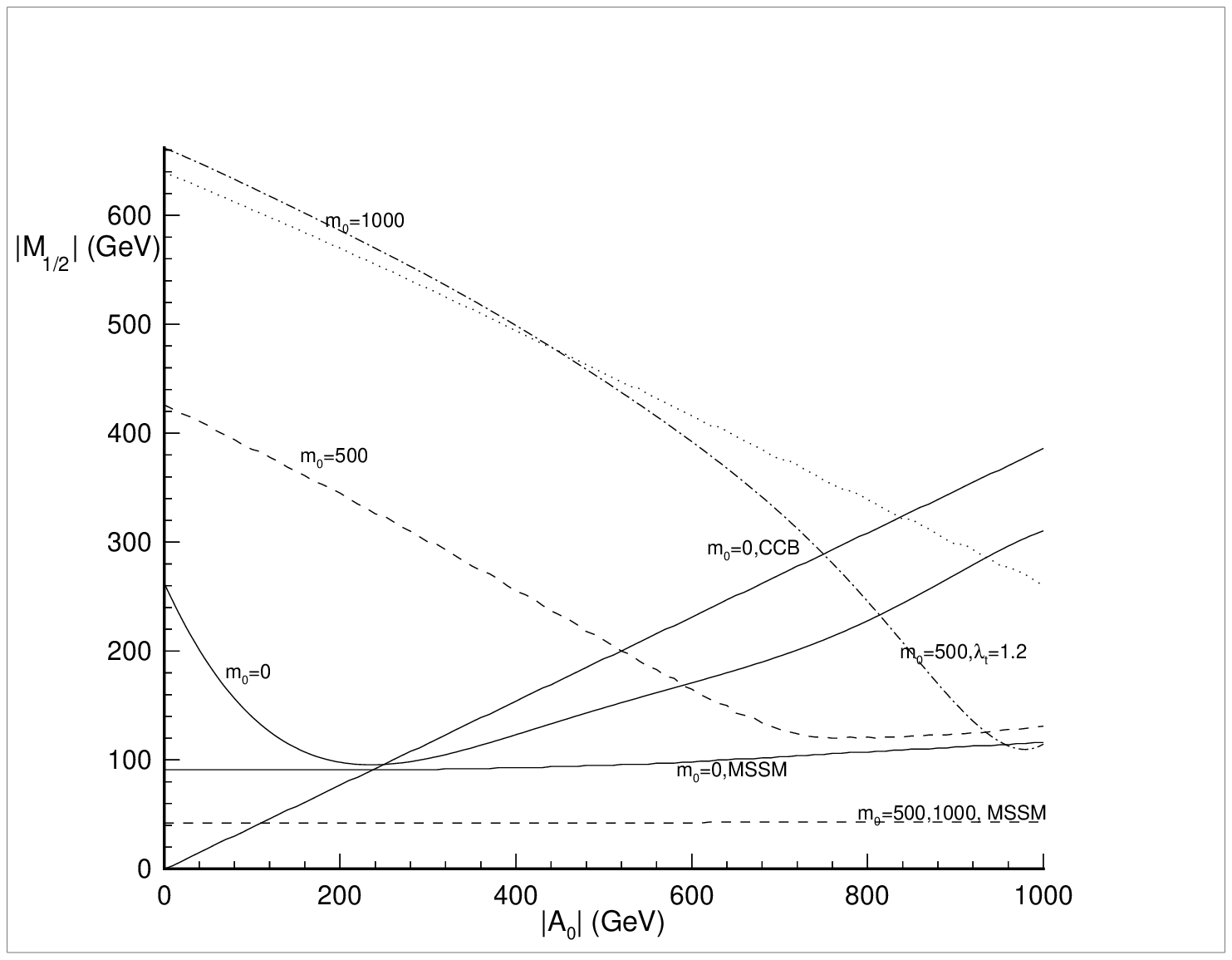 hoffset=-150 voffset=-50
hscale=47 vscale=47}{3cm}{3cm}
\PSbox{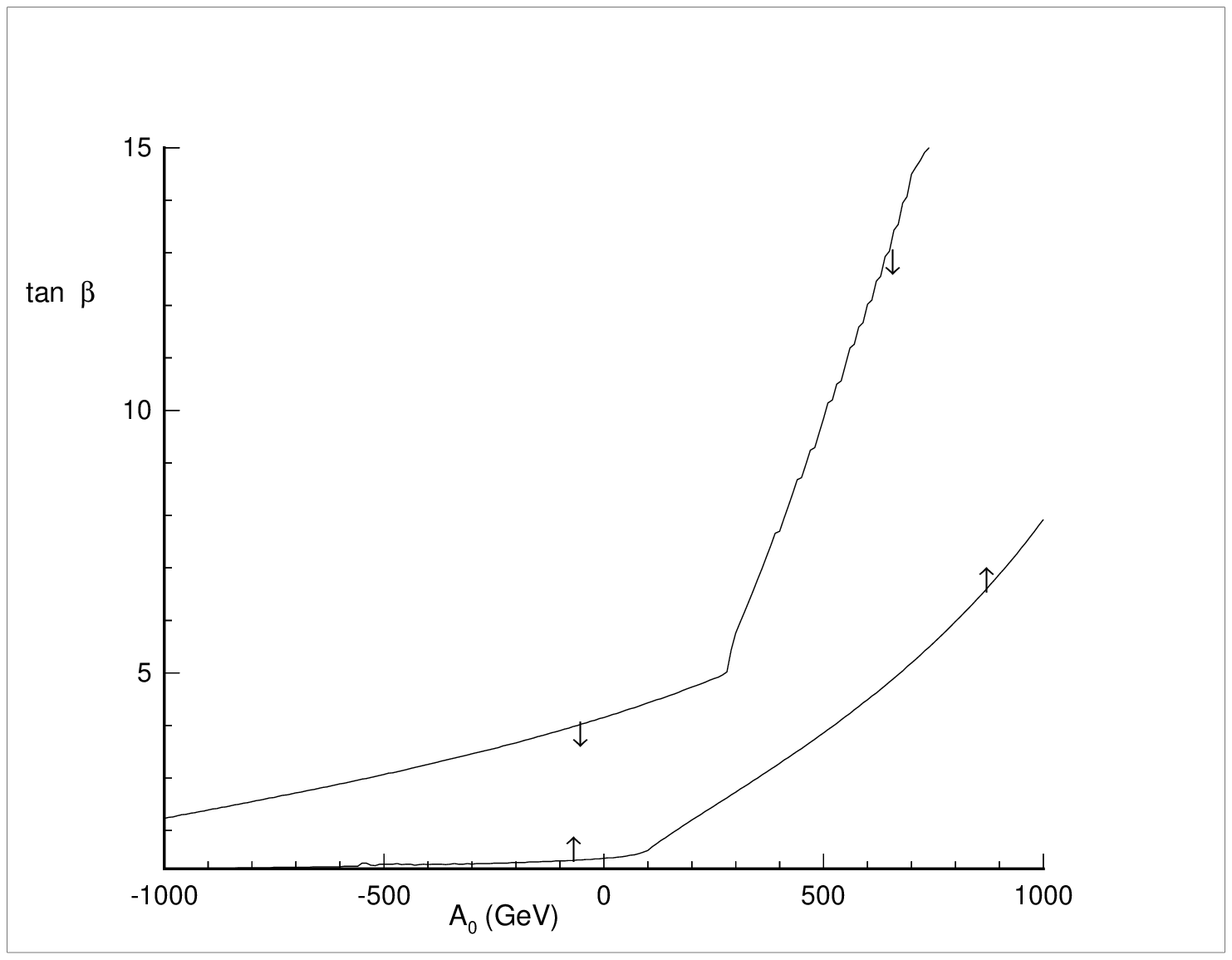 hoffset=-40 voffset=-50
hscale=47 vscale=47}{3cm}{3cm}
\end{center}
\vspace*{1cm}

\caption{The favored parameter region for the presented model.
The figure on the left displays the region of 
$M_{1/2}-A_0$ space where the equation 
for $\tan \beta$ of Ref.$^4$ has a solution for three
different values of $m_0$, with $\lambda_t=0.8$. 
The allowed region for $m_0=0$ is above the solid line, 
for $m_0=500$ GeV above the dashed line and for $m_0=1000$
GeV above the dotted line. The figure on the right displays
the values of the possible solutions of the equation for $\tan
\beta$, for varying $M_{1/2}$, $m_0=0$ and $\lambda_t=0.8$
is fixed in this figure.}
\end{figure}

\newpage

\section*{Acknowledgments} I thank my collaborators Lisa Randall
and Zurab Berezhiani. This work was supported in part by the
US Department of Energy under Cooperative Agreement 
\# DE-FC02-94ER40818.

\section*{Note Added} After this talk has been presented at DPF'96
there has been an interesting new proposal by G.~Dvali and S.~Pokorski
to use the anomalous $U(1)$ symmetry to enforce the accidental
global $SU(6)\times SU(6)$ symmetry~\cite{Dvali}.

\end{document}